# Making thermodynamic functions of nanosystems intensive.

## A M Nassimi[1] and G A Parsafar

Department of Chemistry and Nanotechnology Research Center, Sharif University of Technology, Tehran, 11365-9516, Iran.

Email: ali.nassimi@utoronto.ca and parsafar@sharif.edu

**Abstract**
The potential energy of interaction among particles in many systems is proportional to $r^{-\alpha}$. It has been argued that in systems for which $\alpha < d$, we encounter nonextensive (nonintensive) thermodynamic functions, where $d$ is the space dimension. A scaling parameter, $\tilde{N}$, has been introduced to make the nonextensive (nonintensive) thermodynamic functions of such systems extensive (intensive). Our simulation results show that this parameter is not capable of making the thermodynamic functions of a nanosystem extensive (intensive). Here we have presented a theoretical justification for $\tilde{N}$. Then we have generalized this scaling parameter to be capable of making the nonextensive (nonintensive) thermodynamic functions of nanosystems extensive (intensive). This generalized parameter is proportional to the potential energy per particle at zero temperature.

## 1. Introduction

Ordinary thermodynamics is valid when the number of particles ($N$) in the system goes to infinity. When we have few particles in the system, the surface, rotation and other effects prevent the ordinary thermodynamics from being valid. The first work in this regard was carried out by Hill about five decades ago. He used ordinary thermodynamic relations and added a correction term for each extra effect appearing in small systems [1]. Another approach to nanothermodynamics is to focus on the fluctuations of thermodynamic functions, since fluctuations are not negligible in nanosystems. It has been shown that averaging the fluctuating quantities yields to Tsallis statistics for nanosystems [2,3].

Here we present a method for making the thermodynamic functions of nanosystems extensive. If the interaction between two particles is (at least at long distances) of the form $r^{-\alpha}$, then by considering a fixed particle density and approximating the particle density as continuous, the potential energy ($U$) per particle is of the form

$$\frac{U}{N} \propto \int_1^\infty dr\, r^{d-1} r^{-\alpha}, \qquad (1)$$

---

[1] Current address: Department of Chemistry, University of Toronto, Toronto, ON, M5S 3H6, Canada.

where $d$ is the space dimension. This integral converges when $\alpha > d$; otherwise, it diverges to infinity. Thus in cases where $\alpha < d$, we must take care of the finite dimension of the system and write the relation (1) in the form

$$\frac{U}{N} \propto \int_{1}^{N^{1/d}} dr\, r^{d-1} r^{-\alpha} \quad , \tag{2}$$

where $N^{1/d}$ represents the linear size of the system. According to the relation (2), the energy per particle depends on the number of particles in the system ($N$). This is contrary to what we expect from classical thermodynamics, i.e., energy is an extensive thermodynamic function [4]. We shall refer to the situation $\alpha < d$ as long range interaction. It is claimed that such systems must be investigated using nonextensive statistical mechanics (Tsallis statistics) [5].

In 1995, Jund et al. introduce a parameter-the value of the integral in relation (2) multiplied by $d$-to make $U$ extensive in a system containing long range interactions [6]. This parameter is denoted by $N^*$, i.e.

$$N^* \equiv \frac{N^{1-\alpha/d} - 1}{1 - \alpha/d}. \tag{3}$$

In the limit $N \to \infty$, we will have

$$N^* \approx \begin{cases} \dfrac{1}{\alpha/d - 1} & \text{if } \alpha/d > 1, \\ \ln N & \text{if } \alpha/d = 1, \\ \dfrac{N^{1-\alpha/d}}{1 - \alpha/d} & \text{if } 0 \leq \alpha/d < 1. \end{cases} \tag{4}$$

Then Tsallis inspected some systems containing long-range interactions, i.e., the $d$-dimensional Bravais-lattice Ising ferromagnet and the system studied by Jund et al. [6]. Tsallis propose that, for making all thermodynamic functions appropriately intensive or extensive, we must divide all functions having the dimension of energy and all functions which are defined through a partial derivative of energy by $N^*$, [7].

Then we can define the reduced energy, enthalpy, free energy, thermodynamic potential, temperature and pressure respectively via the relations $E^* = E/N^*$, $H^* = H/N^*$, $A^* = A/N^*$, $G^* = G/N^*$, $T^* = T/N^*$, $p^* = p/N^*$. By using these newly defined functions we regain all usual extensive (intensive) properties of ordinary thermodynamics. Also, each thermodynamic function is a function of these new variables to be well behaved. We can easily see that $(\frac{\partial E^*}{\partial S})_{N,V} = T^*$ and $(\frac{\partial E^*}{\partial V})_{N,S} = -p^*$. So defining $H = E - V(\frac{\partial E}{\partial V})_{S,N}$, we can get $H^* = E^* + p^* V$, and similar relations for the free energy. This means that the Legendre transformation structure of thermodynamics is not affected with this scaling. Here we note that intensive functions in the standard structure of thermodynamics and functions with the energy dimension must be scaled by $N^*$, then all thermodynamic functions are functions of these scaled quantities [7].

In 1996, Cannas and Tamarit used this scaling parameter to generalize the Currie-Weiss model for an Ising ferromagnet [8]. In the same year, Grigera carried out some MD simulations on a lattice with a generalized Lennard-Jones (LJ) potential, i.e., $V_{ij} = C_{12} r^{-12} - C_\alpha r^{-\alpha}$. He showed that by plotting $E/NN^*$ at constant $T^*$, one can get a horizontal line. He also showed that the potential energy is a very weak function of the scaled temperature [9].

In 1997, Sampaio et al. used this scaling for thermodynamics of magnetic systems. They obtained magnetization curves of a two-dimensional classical Ising-model, including an interaction potential proportional to $r^{-\alpha}$. They deduced that, the appropriate form of the equation of state for magnetic systems is $M/N = m(T^*, H^*)$ [10].

In 1997, Cannas and de Magalhaes studied the one-dimensional potts model with long-range interactions. They verified that, we must scale the critical temperature by *N\** to get a constant critical temperature for different values of $\alpha$ [11].

In 1999, Salazar and Toral studied the one-dimensional Ising-model with long-range interactions in the context of Tsallis statistics. They used the aforementioned scaling as a starting point. It is only valid for *q* = 1, where *q* is the entropy index introduced in Tsallis Statistics [5]. They generalize the Monte-Carlo simulation to $q \neq 1$, and find different scaling for the cases *q* < 1, *q* = 1, and *q* > 1 [12]. In 1999, Curilef and Tsallis investigated a LJ-like fluid. They calculated the liquid vapor critical point as a function of $\alpha$ and *N*, i.e., $T_c(\alpha,N)$ and $p_c(\alpha,N)$. They shown that in order to avoid getting a negative pressure we must use *T\** instead of *T* [13].

In 1999, Tsallis argued that in the limit $\alpha/d \to \infty$, which means very short ranged interactions (the nearest neighbor), we have $N^* \to 0$ which is an unphysical scaling. Thus it is better to define $\tilde{N}$ as

$$\tilde{N} \equiv N^* + 1 = \frac{N^{1-\alpha/d} - \alpha/d}{1 - \alpha/d}. \tag{5}$$

He also mentioned that, $\tilde{N}$ characterizes the effective number of neighbors that can be associated with a given particle [14].

In 2005, Abe and Rajagopal defined $\lambda = (2 - \alpha/d)^{\frac{-1}{1-\alpha/d}}$, then by implicitly redefining temperature, pressure and chemical potential respectively via $T \equiv \frac{1}{\lambda}\frac{\partial U}{\partial S}$, $p \equiv -\frac{1}{\lambda}\frac{\partial U}{\partial V}$ and $\mu \equiv \frac{1}{\lambda}\frac{\partial U}{\partial N}$, they proved the Euler relation, i.e. $U = TS - pV + HM + \mu N$. Note that the definitions of *T*, *p* and $\mu$ are $\alpha/d$ dependent. They concluded that, the Gibbs-Duhem relation must be valid for these systems. Therefore, $\mu$, *T*, *p* and *H* must be scaled with *N\**. This proves *N\** scaling from a classical thermodynamics point of view [15].

This scaling factor has been satisfactorily used for various systems [16, 17, 18, 19, 20, 21, 22, 23]. In this work, we first justify the use of this scaling factor and then argue that it can not make the thermodynamic functions of nanosystems extensive (intensive). By defining a new scaling factor, i.e., *N'*, which has the same physical interpretation but is computed differently, we will extend this method to the realm of nanosystems, in order to make nanosystem thermodynamic functions extensive (intensive).

## 2. Theory

We have assumed that the interaction energy among particles in the system is of the form $r^{-\alpha}$. Therefore, according to the relation (2) and the definition (3) the potential energy of the system is proportional to *N\**. Now, we consider the parameter *N'*, which in the thermodynamic limit (TL), is equal to $N^* = \tilde{N}$ and is defined in section 4 for nanosystems. The virial theorem in statistical mechanics says [24]

$$(2-\alpha)\langle K \rangle = -\alpha \langle E \rangle + 3pV, \tag{6}$$

where *K* is the kinetic energy of the system. Thus, the dependence of *U, K, E* and *pV* on *N* must be the same. This means that when we define a scaling parameter *N'* such that it scales *U* to an extensive parameter, *N'* must do the same for *K* and *E* as well.

Enthalpy is defined as $H \equiv U + pV$; thus, *N'* is the scaling factor making enthalpy extensive. Now, let us consider pressure defined as $p = -(\frac{\partial E}{\partial V})_{N,S}$; dividing both sides by *N'*, we get

$$\frac{P}{N'} = -(\frac{\partial (E/N')}{\partial V})_{N,S} . \qquad (7)$$

The numerator of the right hand side of the relation (7) is extensive. considering the concept of volume in solids and liquids, we expect the denominator to be extensive; therefore $p$ must scale by $N'$ to be intensive.

According to the kinetic theory, for a substance whose molecules do not possess internal degrees of freedom, temperature is defined through [25]

$$K = \frac{1}{2} m \langle (v - \langle v \rangle)^2 \rangle = \frac{3 N k_B T}{2} , \qquad (8)$$

where $v$ is a particle's velocity, and $\langle \rangle$ represents an average over all particles in the system. Having internal degrees of freedom changes the constant of proportionality. Since $K$ scales by $NN'$, considering equation (8), $T$ scales with $N'$.

Helmholtz free energy is defined through the equation $A \equiv E - TS$, so $A/N'$ is extensive. We have shown that $T/N'$ is intensive, so $S$ is extensive. With a similar argument, we can show that Gibbs free energy ($G$) scales by $NN'$.

In the case of systems with long range interactions, we have either a lattice with long range interacting rotators (spins), or a LJ-like fluid. In the first case, the only potential present in the system is of the form $r^{-\alpha}$; in the second case, the prevailing term is of the form $r^{-\alpha}$. Thus, we can use the result of the virial theorem, equation (6). But in the case of nanosystems, e.g., for the common two body interaction potentials, we encounter two terms of the form $r^{-\alpha}$, so the virial theorem doesn't give any result. Imagine that the kinetic and the potential energy scale with different powers of $N$, then if the kinetic energy scales with a larger power of $N$, in the TL we will have a potential energy negligible compared to the kinetic energy, which practically means no interaction. If the potential energy scales with a larger power of $N$, then in the TL the kinetic energy will be negligible compared to the potential energy, which means the system is frozen. Since there exist interactions and the system is not frozen, potential and kinetic energy must scale in the same way with $N$, even in the absence of the virial theorem.

## 3. Scaling Parameter

In the previous discussion, we defined the scaling factor as the average number of interactions per particle in the continuum approximation. Switching to nanosystems, we can not use the continuum approximation and compute the average by integration.

As can be seen in Figure 1, which represents the internal and potential energy per particle versus $N$, nonextensivity in nanosystems strongly depends on the surface size. The average interaction energy for a surface particle is different from this average for a particle in bulk. Changing the shape or size of the system changes the fraction of the surface particles. Thus, the scaling parameters $N^*$ and $\tilde{N}$, which do not depend on the shape of the system, are not relevant for nanosystems, e.g., scaling the results of our simulations (section 4) with these parameters yields no improvement in their linear correlation coefficient. Therefore, defining an analytical form for the scaling parameter $N'$ in nanosystems seems to be impossible.

The concept of energy per particle is an exact concept in some systems, e.g., ideal gases, phonons and photons; it is also a useful concept in the realm of systems containing short-range interactions, but in the realm of systems containing long-range interactions or nanosystems, we must instead use the concept of energy per number of effective interactions, which is proportional to $N$ multiplied by $N'$. Here, by considering the particles static in their equilibrium position (classical system in zero temperature), we want to compute the average of the interactions of each particle with other particles denoting this by $N'$. In section 4, how fast the strength of interactions falls with distance is shown. After the distance related to the fifth neighbor, the strength of the interactions are negligible; therefore, we just considered the first fifth layers of neighbors in computing $N'$.

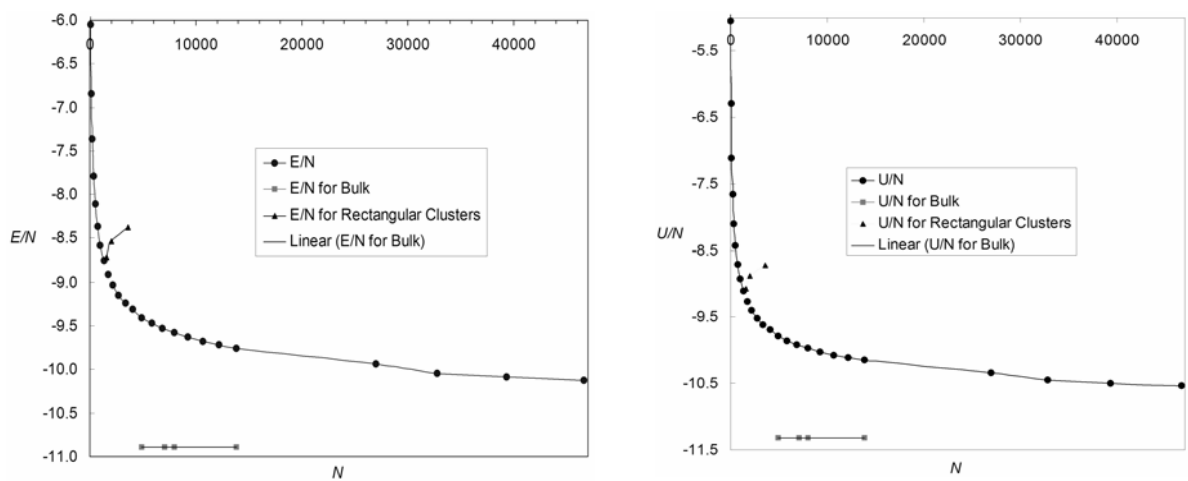

a)　　　　　　　　　　　　　　　　　　　b)
**Figure 1.** Calculated internal energy (a) and potential energy (b) per atom for Kr clusters (in kJ/mol). Note the strong dependence of *E/N* on the cluster surface.

**4. Molecular Dynamic Simulation**
The simulation was performed by using the DL_POLY code [26]. An atomic cluster containing noble gas atoms can well be described by the LJ potential. We have performed our simulation on a FCC lattice of Kr atoms with LJ parameters, $\sigma = 3.827$Å and $\varepsilon/k_B = 164.0$ K [27]. Due to software limitations, we have performed our simulations at zero pressure. Then, we had to use low temperatures to prevent the cluster from sublimation. We have used the canonical ensemble. The Evans method was used for fixing the temperature of the simulation cell.

Since the scaled temperature must be the same for all the simulated systems, we consider the value of 2.1419 as the reduced temperature, and multiply *N'* by this value in order to evaluate the simulation temperature for that system. We have plotted the graph of temperature versus simulation step and found it to have large and directional fluctuations until step 300, so we consider the first 500 steps of each simulation for equilibration and evaluate the average by the data from steps 500 to 10000. The time between simulation steps was 0.002 ps; One of the systems was simulated by a time step of 0.001 ps which yielded the same results. The cutoff for the van der Waals forces was considered to be 16Å for the smaller systems, and larger for larger systems. Since the nearest neighbor separation in this lattice is about 4Å, and the prevailing term in LJ potential is the $(\sigma/r)^6$ term, so the interaction strength is about $\frac{1}{4}^6 = 1/4096$ of its initial value. When instead of two atoms we have a lattice of atoms, the equilibrium separation between LJ interacting atoms will be 0.971 times that of two separate atoms [28]. Therefore, the nearest neighbor separation in this lattice is 4.171Å.

The computed scaling parameters and simulation temperatures, as well as the results for the *E, U,* and their scaled values for atomic Kr clusters are reported in Table 1. Since, *p* is zero for these clusters, *H* is equal to *E* in the clusters. The computed values of scaling parameters and simulation temperatures together with the values of *E, U* and *H* are given in Table 2 for similar systems simulated with the periodic boundary condition. The energy and potential energy per particle are sketched in Figure 1 to show nonextensivity in such systems. The points represented with a circle are atomic clusters with an equal number of atoms along each axe of the crystal. These systems contain $n^3$ atoms where *n* = 2, …, 24, 30, 32, 34, and 36. In order to find the thermodynamic functions in the TL, we have also simulated three systems with an equal number of atoms along the three axes and, a system with (14, 21, 24) atoms along its axes are simulated with the periodic boundary condition. They correspond to four points constituting a line at the bottom of Figure 1. Another three systems containing (7, 15, 15), (5, 20, 20) and (4, 30, 30) atoms along each axe have been simulated to show the effects due to shape and surface. Figure 1 clearly shows that these systems do not fit with the trend of the previous systems. Figure 2 shows the graphs of $E/N'$ and $U/N'$ versus $N$. The success of this method for making thermodynamic functions extensive is reflected in the value of 1.0000 for the linear correlation coefficient in these graphs.

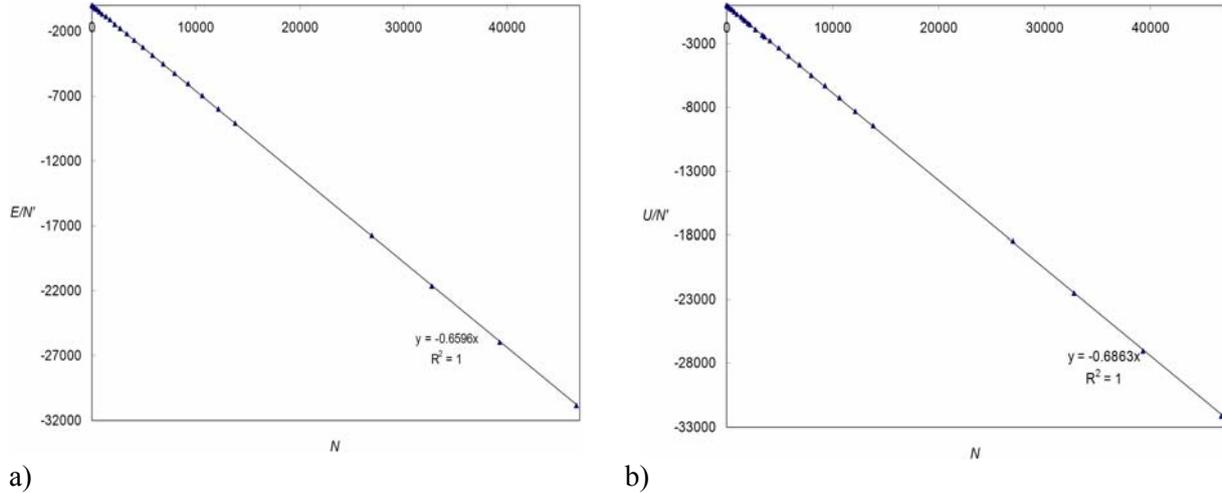

a) b)

**Figure 2.** calculated internal energy (a) and potential energy (b) for Kr clusters scaled with $N'$ for the reduced temperature $T/N' = 2.142$, note the linear correlation coefficient of 1.0000.

**5. Conclusion**
In this work, we have justified the scaling method that was used to make extensive (intensive) thermodynamic functions in systems containing long range interactions [6-23], which are nonextensive (nonintensive) on their own. We have argued that by going to limited systems, i.e., where we do not reach the TL, thermodynamic functions are again nonextensive. (See figure 1.) Our justification shows that this nonextensivity yields nonintensivity. The scaling parameter for doing this must depend on the shape of the system. Thus, in the case of these systems, we discarded $N^*$ and $\tilde{N}$, and defined $N'$, which has the same interpretation but must be evaluated without using the continuum approximation, i.e., integration. Finally, we have performed an MD simulation on some atomic clusters of Kr; the results are presented in Tables 1 and 2. We have scaled the internal and potential energy by the parameter $N'$; The results are represented in Figure 2, with the linear correlation coefficient of 1.0000 for both graphs.

This scaling is universal for nanocrystals since there is no assumption beyond that the system will exhibit normal behavior in TL; But, going to liquids the particles will not have definite positions relative to each other, so we can not evaluate $N'$ in this manner. If the range of interactions is short enough for one or two shells of neighbors to be enough for evaluating the $N'$, we can determine the dependence of the $N'$ on the portion of the particles present on the surface. But this depends on the number of the nearest neighbors for each group of particles (surface and bulk), which in turn depends on the lattice structure.

Doing some simulation capable of computing entropy and free energy remains for the future, to check the validity of this scaling on such systems, and also, to check the scaling of the thermodynamic functions of nanomagnetic systems.

This method makes thermodynamic functions artificially extensive to broaden the range of the applicability of the ordinary thermodynamics to systems containing long-range interactions and nanosystems. But some guesses have been made that such systems must be described by nonextensive thermodynamics [5], systems containing long rang interactions with a value of entropy index $q$ that is a function of $\alpha$ and $d$ and nanosystems with a value of $q$ that is a function of $\alpha$, $d$, and $N$ [6]. The result of this work shows that the mentioned $q$ for nanosystems must also be a function of the system shape (to account for surface effects). Investigation into this guess also remains for the future.

**Acknowledgement**
We acknowledge Iranian Nanotechnology Initiative and Sharif University of Technology for financial support. We acknowledge A. Sheikhan for useful discussion.

**Table 1.** the calculated thermodynamic functions obtained from the MD simulation for krypton clusters, with given number of atoms ($N$) and absolute temperature ($T$).

| $N$ | $N'$ | $T/K$ | $E$ (kJ/mol) | $-E/N'$ | $-U$ (kJ/mol) | $U/N'$ |
|---|---|---|---|---|---|---|
| 8 | 4.669 | 10.00 | 2.45E+01 | 5.257 | 25.29 | -5.417 |
| 27 | 7.596 | 16.27 | 1.31E+02 | 17.28 | 136.4 | -17.95 |
| 64 | 9.380 | 20.09 | 3.87E+02 | 41.28 | 402.7 | -42.94 |
| 125 | 10.56 | 22.62 | 8.55E+02 | 81.01 | 889.7 | -84.25 |
| 216 | 11.40 | 24.41 | 1.59E+03 | 139.4 | 1653 | -145.1 |
| 343 | 12.02 | 25.74 | 2.67E+03 | 222.4 | 2780 | -231.4 |
| 512 | 12.50 | 26.76 | 4.15E+03 | 332.1 | 4318 | -345.6 |
| 729 | 12.88 | 27.58 | 6.10E+03 | 473.7 | 6347 | -492.9 |
| 1,000 | 13.19 | 28.24 | 8.58E+03 | 650.7 | 8927 | -677.0 |
| 1,331 | 13.44 | 28.79 | 1.17E+04 | 867.0 | 12126 | -902.1 |
| 1,575 | 13.40 | 28.70 | 1.37E+04 | 1024 | 14289 | -1066 |
| 1,728 | 13.66 | 29.26 | 1.54E+04 | 1127 | 16015 | -1172 |
| 2,000 | 13.14 | 28.14 | 1.71E+04 | 1299 | 17762 | -1352 |
| 2,197 | 13.84 | 29.65 | 1.99E+04 | 1434 | 20653 | -1492 |
| 2,744 | 14.00 | 29.99 | 2.51E+04 | 1792 | 26111 | -1865 |
| 3,375 | 14.14 | 30.29 | 3.12E+04 | 2206 | 32456 | -2295 |
| 3,600 | 12.93 | 27.70 | 3.02E+04 | 2333 | 31408 | -2429 |
| 4,096 | 14.26 | 30.55 | 3.82E+04 | 2675 | 39692 | -2783 |
| 4,913 | 14.37 | 30.78 | 4.62E+04 | 3216 | 48083 | -3345 |
| 5,832 | 14.47 | 30.99 | 5.52E+04 | 3815 | 57457 | -3970 |
| 6,859 | 14.56 | 31.18 | 6.53E+04 | 4488 | 67996 | -4671 |
| 8,000 | 14.64 | 31.35 | 7.66E+04 | 5236 | 79764 | -5450 |
| 9,261 | 14.71 | 31.50 | 8.92E+04 | 6063 | 92816 | -6311 |
| 10,648 | 14.77 | 31.64 | 1.03E+05 | 6973 | 107210 | -7257 |
| 12,167 | 14.83 | 31.77 | 1.18E+05 | 7969 | 123020 | -8294 |
| 13,824 | 14.89 | 31.89 | 1.35E+05 | 9056 | 140310 | -9425 |
| 27,000 | 15.14 | 32.43 | 2.68E+05 | 17712 | 279090 | -18433 |
| 32,768 | 15.20 | 32.57 | 3.29E+05 | 21649 | 342470 | -22524 |
| 39,304 | 15.26 | 32.69 | 3.96E+05 | 25976 | 412440 | -27026 |
| 46,656 | 15.31 | 32.79 | 4.72E+05 | 30845 | 491360 | -32092 |

**Table 2.** the calculated energy ($E$), potential energy ($U$) and enthalpy ($H$) obtained for systems simulated with the periodic boundary condition, with the given number of atoms ($N$).

| $N$ | $E$ (kJ/mol) | $U$ (kJ/mol) | $H$ (kJ/mol) |
|---|---|---|---|
| 4,913 | -5.35E+04 | -5.56E+04 | -4.09E+04 |
| 7,056 | -7.68E+04 | -7.99E+04 | -5.88E+04 |
| 8,000 | -8.71E+04 | -9.06E+04 | -6.66E+04 |
| 13,824 | -1.51E+05 | -1.56E+05 | -1.15E+05 |

$N'$ is equal to 16.182 and $T$ is equal to 34.660.